\documentclass[jap,showpacs,twocolumn,groupedaddress]{revtex4}

\usepackage{graphicx,amsmath,amssymb}
\usepackage[utf8]{inputenc}
\usepackage[T1]{fontenc}

\newcommand{\fig}[4][h]{\begin{figure}[#1]\begin{center}\includegraphics[scale=#2]{figures/#3.pdf}\vspace{-0.25 cm}\caption{#4}\label{fig:#3}\end{center}\end{figure}}

\renewcommand{\vec}[1]{\mathbf{#1}}
\newcommand{\ket}[1]{\ensuremath{\left|{#1}\right\rangle}}
\newcommand{\Rb}[1]{\ensuremath{^{#1}}\text{Rb}}
\newcommand{\hf}[2]{\ket{#1, \; #2}}
\newcommand{\etal}[0]{\emph{et al.}}

\begin{document}

\title{Bosenova and three-body loss in a \Rb{85} Bose-Einstein condensate}

\author{P.~A. Altin, G.~R. Dennis, G.~D. McDonald, D. D\"oring, J.~E. Debs,\linebreak J.~D. Close, C.~M. Savage and N.~P. Robins}
\affiliation{Department of Quantum Science, Research School of Physics and Engineering, Australian National University, ACT 0200, Australia}
\email{paul.altin@anu.edu.au}
\homepage{http://atomlaser.anu.edu.au/}

\date{\today}

\begin{abstract}

Collapsing Bose-Einstein condensates are rich and complex quantum systems for which quantitative explanation by simple models has proved elusive. We present new experimental data on the collapse of high density \Rb{85} condensates with attractive interactions and find quantitative agreement with the predictions of the Gross-Pitaevskii equation. The collapse data and measurements of the decay of atoms from our condensates allow us to put new limits on the value of the \Rb{85} three-body loss coefficient $K_3$ at small positive and negative scattering lengths.

\end{abstract}


\maketitle

While most experiments with dilute gas Bose-Einstein condensates have employed atomic species with repulsive interactions, it has long been known that interesting and exotic physics manifests in attracting condensates. These include macroscopic quantum tunnelling \cite{Ueda:1998}, the formation of soliton trains and vortex rings \cite{Cornish:2006,Lahaye:2008}, and the violent collapse and explosion known as the `bosenova' \cite{Saito:2001,Saito:2001a,Donley:2001}. The first evidence for the collapse of attracting Bose-Einstein condensates was found by Sackett and coworkers, who analysed the thermal equilibration of a sample of $^7$Li atoms with negative scattering length that was cooled below the critical temperature \cite{Sackett:1999}. Soon after this work, condensate collapse was directly observed in pioneering experiments at JILA \cite{Donley:2001}, which revealed a host of interesting dynamics and prompted a surge of theoretical interest \cite{Saito:2002,Adhikari:2002,Santos:2002,Savage:2003,Milstein:2003,Bao:2004,Adhikari:2004,Wuster:2005}. More recently, the collapse of dipolar chromium BECs has been observed, displaying the striking $d$-wave symmetry of long-range dipole-dipole interactions in excellent agreement with theory \cite{Lahaye:2008}.

However, while initial mean-field analysis of the JILA experiment using the Gross-Pitaevskii (GP) equation was able to qualitatively account for most of the experimental observations, including the formation of atomic `bursts' and `jets' \cite{Saito:2002,Milstein:2003,Bao:2004,Adhikari:2004,Wuster:2005}, further investigation exposed a quantitative discrepancy between theory and experiment in the time taken for the condensates to collapse \cite{Savage:2003}. This was especially puzzling as the short time, low density phase of the experiment is exactly where the GP equation should be an excellent approximation. This disagreement, of about 100\%, could not be eliminated by more complex quantum field calculations \cite{Wuster:2005,Wuster:2007}, and has led to the development of competing models for the collapse mechanism \cite{Calzetta:2008}. Yet amid the extensive theoretical work on this phenomenon that has continued in recent years, there has been a notable absence of further experimental data, and the discrepancy between theory and the \Rb{85} experiment remains unresolved.

Here we present the first results on this phenomenon from a new \Rb{85} BEC machine \cite{Altin:2010a}, finding good agreement between the measured collapse times and those predicted by a GP model. Although we use the same atom, our experiment has several important differences from the original JILA work. Most notably, our condensates are confined in a purely optical potential, with a homogeneous magnetic bias field applied to manipulate the interatomic interactions. In addition, we have measured condensate collapses with $4\times10^4$ atoms in a tighter trap, which together result in an initial density over an order of magnitude larger than in Ref. \cite{Donley:2001}. This leads to shorter collapse times and lower values of the critical scattering length, but should not affect the ability of mean-field theory to describe the evolution of the system. It also allows us to investigate three-body recombination rates in a high density regime where they are the dominant source of atom loss.

Our apparatus for producing Bose-Einstein condensates of \Rb{85} with tunable interactions has been described in detail elsewhere \cite{Altin:2010a}. In brief, we employ sympathetic cooling using \Rb{87} as a refrigerant, initially in a quadrupole-Ioffe configuration magnetic trap and subsequently in a weak, large-volume crossed optical dipole trap. During the final evaporation, a magnetic bias field of 167\,G is applied to reduce losses due to two-body inelastic collisions \cite{Roberts:2000,Altin:2010}. We can create condensates of up to $10^5$ \Rb{85} \hf{F=2}{m_F=-2} atoms with a thermal fraction below 10\% in a trap with harmonic oscillation frequencies $\omega_{x,y,z} = 2\pi\times\{53,22,27\}$\,Hz. Condensates form at scattering lengths between $a=+50a_0$ and $a=+200 a_0$, where $a_0$ is the Bohr radius. We determine the scattering length from the applied magnetic bias field using the known parameters of the 155\,G \Rb{85} Feshbach resonance \cite{Claussen:2003}. The field is calibrated by addressing radiofrequency transitions between the $m_F$ sublevels of the $F=2$ manifold; the transition frequency is related to the magnetic field strength by the Breit-Rabi equation \cite{Breit:1931}. The magnetic field can be determined in this way to within 5\,mG, which near the zero crossing of the scattering length corresponds to an uncertainty in $a$ of $\pm0.2a_0$.

To observe condensate collapse, we follow the procedure of Donley \etal, tuning the atomic interactions using the Feshbach bias magnetic field as shown in Figure \ref{fig:collapse}(a). First, the scattering length is ramped smoothly from $a=+89.1a_0$ to $a_\text{init}=+0.2a_0$ over 100\,ms to produce a near-ideal, noninteracting gas. The magnetic field is then increased suddenly ($<100\,\mu$s) to a value at which the interactions are attractive $a_\text{collapse}<0$, and held there for a time $\tau$ before the trap is switched off and the scattering length simultaneously increased to $a=+50a_0$. The condensate is allowed to expand ballistically at this value for 15\,ms, after which the magnetic field is switched off, changing the scattering length to $a_\text{bg} = -443a_0$. Following a futher 5\,ms of free evolution, the number of atoms present is determined by absorption imaging.

The number of atoms remaining as a function of evolution time at $a_\text{collapse}=-20a_0$ is shown in Figure \ref{fig:collapse}(b). In agreement with the original work of Donley \etal, we observe a sudden and delayed onset of atom loss. This is explained by density-dependent three-body recombination; when the interactions are made attractive, the condensate begins to contract slowly and its peak density $n_0$ increases, although not enough to cause significant three-body loss. As the condensate shrinks, however, the contraction accelerates, resulting eventually in a sudden implosion which increases the density by several orders of magnitude. This induces significant recombination losses (the loss rate scales with $n^3$) which ultimately halt the growth in density. The subsequent dynamics include further sporadic local implosions, which effect decay of the atom number in an approximately exponential form. We have observed remnant clouds surviving long after the collapse which contain several times the critical number of atoms $N_\text{cr} \simeq 0.6\,a_\text{ho}/|a|$, where $a_\text{ho} = \sqrt{\hbar/m\omega_\text{ho}}$ is the harmonic oscillator length \cite{Ruprecht:1995} (the critical number for a condensate with $a=-20a_0$ in our trap is $N_\text{cr} \simeq 1200$; c.f. Figure \ref{fig:collapse}). Such configurations have been shown to achieve stability through the formation of mutually repelling bright solitons \cite{Cornish:2006}.

\fig[t]{0.42}{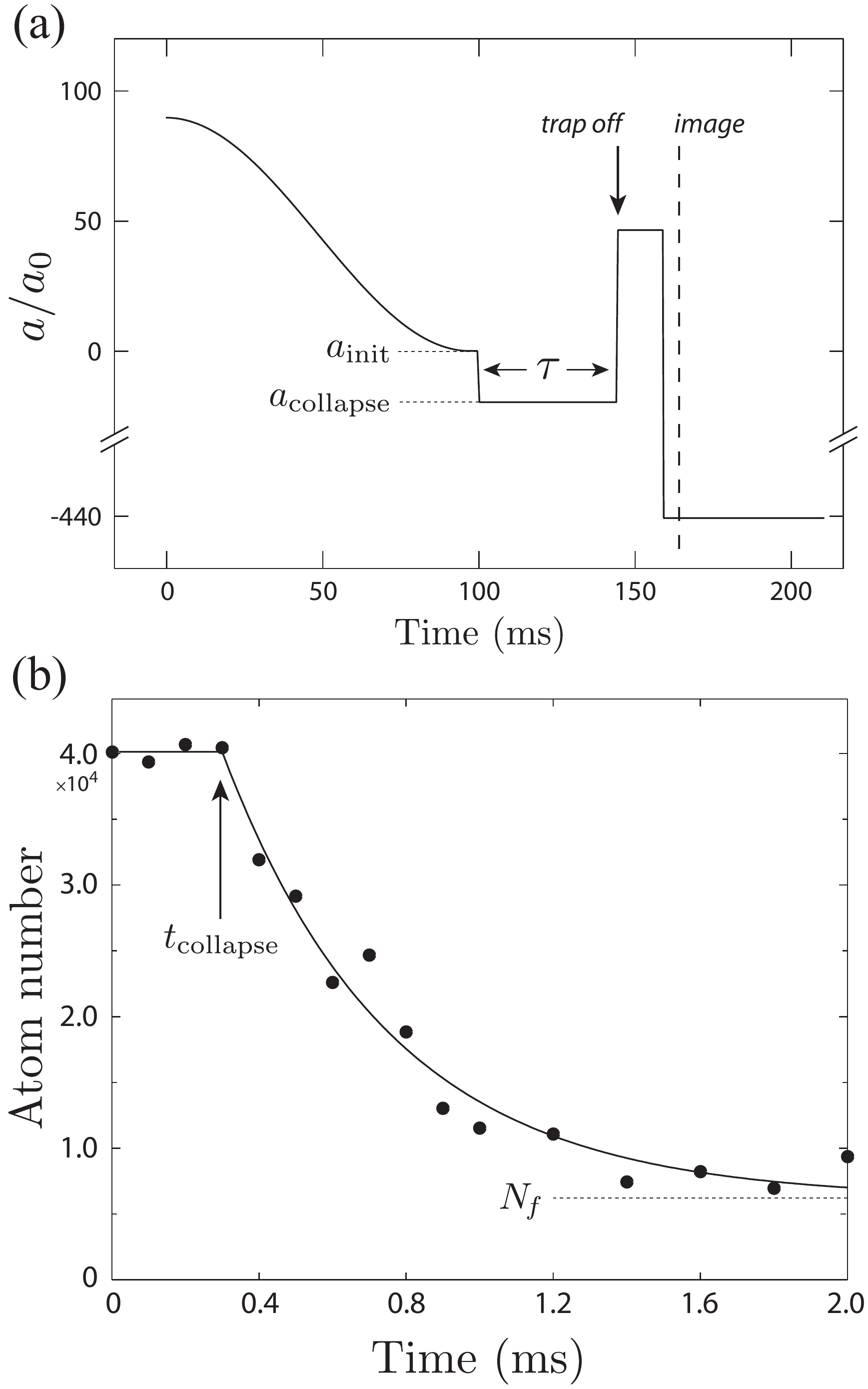}{(a) Manipulation of the scattering length to induce and observe condensate collapse. After a variable evolution time $\tau$, the atoms are released from the optical trap simultaneously with an increase in $a$ from $a_\text{collapse}$ to $+50a_0$. The cloud is allowed to expand at this value for 15\,ms before the magnetic bias field is switched off. (b) Measured atom number as a function of $\tau$ for $a_\text{collapse}=-20a_0$. The solid line is a fit of the experimental data to equation (\ref{eqn:collapsefit}). The atom number remains approximately constant for a time $t_\text{collapse}$, before a sudden onset of loss due to three-body recombination.}

The discrepancy between the JILA experiment and theoretical models concerns one of the most elemental characteristics of the bosenova: the time for which the atom number remains constant before the first density implosion -- the so-called `collapse time' $t_\text{collapse}$. Although the dynamics after the collapse are predicted to be complex and may exhibit behaviour beyond mean-field effects, the evolution prior to the first implosion should be captured in the mean-field approximation, and is determined almost exclusively by the initial density which is experimentally constrained. In particular, it has been noted that $t_\text{collapse}$ does not depend strongly on the the three-body recombination rate $K_3$ \cite{Saito:2002,Savage:2003}, which is not well-determined in the vicinity of the Feshbach resonance. Despite this, Gross-Pitaevskii (GP) simulations were found to systematically overestimate the collapse time measured in the JILA system by almost 100\%, a discrepancy at the $2\sigma$ level given the experimental uncertainties \cite{Savage:2003}.

As in previous work, we determine the collapse time by fitting plots of the measured atom number versus time to the function
\begin{equation}
N(t) = (N_0 - N_f) \exp\left[-\frac{(t-t_\text{collapse})}{\tau_\text{decay}}\right] + N_f \,,
\label{eqn:collapsefit}
\end{equation}
for $t>t_\text{collapse}$, where $N_0$ and $N_f$ denote the atom number at $t<t_\text{collapse}$ and $t\gg t_\text{collapse}$ respectively. Figure \ref{fig:collapsetimes} shows the collapse time determined in this way as a function of $a_\text{collapse}$ for samples of $N_0=4\times10^4$ atoms. As expected, the collapse time is shorter for larger $|a_\text{collapse}|$, as stronger attraction between the condensate atoms results in more rapid contraction. The data are in qualitative agreement with the original experiment of Ref. \cite{Donley:2001} and later theoretical work.

\fig[b]{0.48}{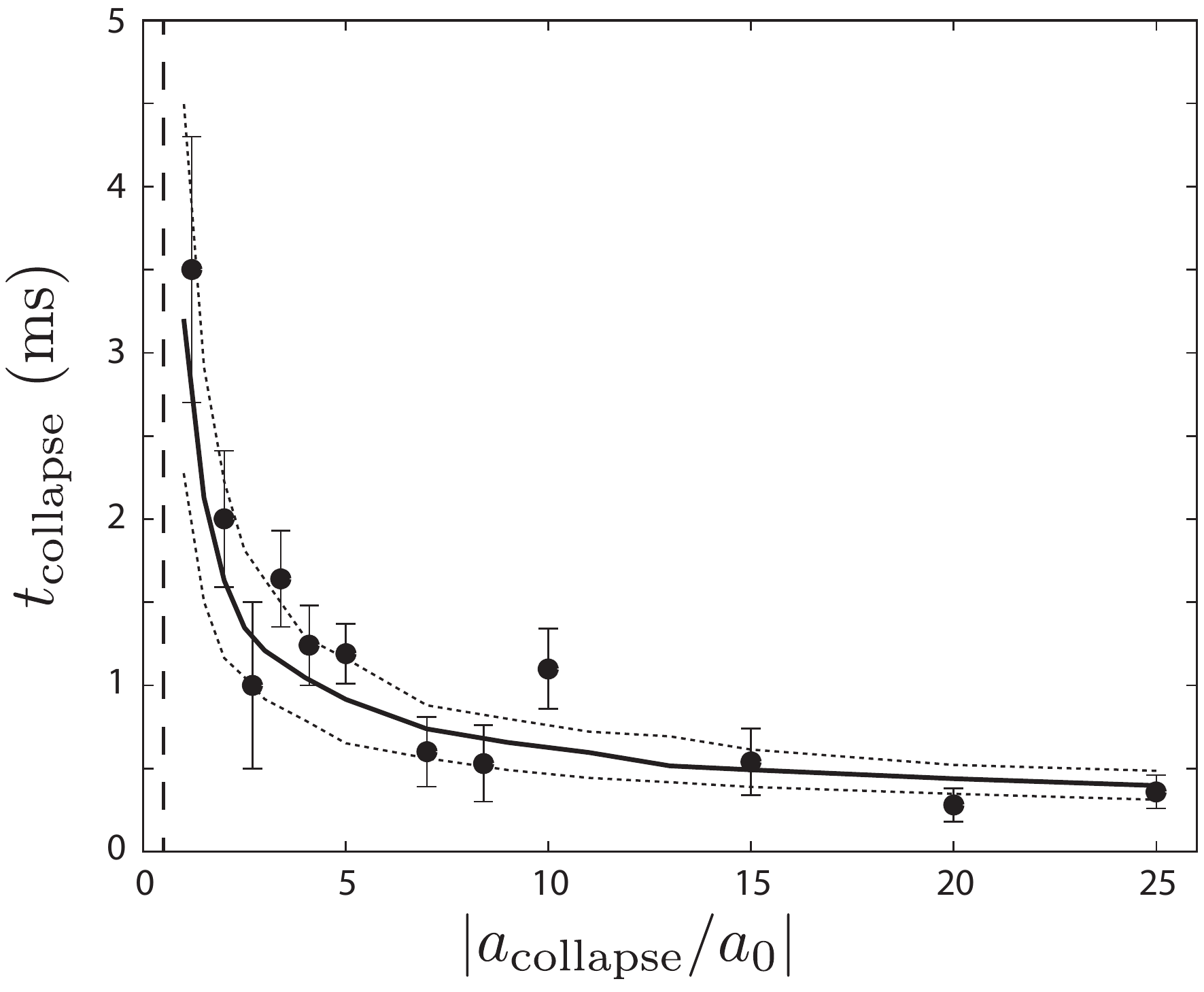}{Collapse times as a function of scattering length for $a_\text{init}=+0.2a_0$ and $N_0 = 4\times10^4$ atoms. The data points represent experimental values from measured decay curves such as that shown in Figure \ref{fig:collapse}(b), with error bars denoting the statistical uncertainty in the fit of equation (\ref{eqn:collapsefit}). The solid line is the result of GP simulations for our experimental parameters, and shows good quantitative agreement with the experimental data. The dotted lines represents the variation in the simulated collapse time due to experimental uncertainties in $a_\text{init}$, $a_\text{collapse}$, $\bar\omega$ and $N_0$. The dashed vertical line shows the critical scattering length for collapse at this atom number, below which the condensate's kinetic energy stabilizes it against implosion.}

To ascertain the ability of mean-field theory to \emph{quantitatively} reproduce our experimental data, we have performed numerical simulations for the parameters of our system using the Gross-Pitaevskii equation for the condensate wave function $\Psi$:
\begin{equation}
i\hbar \frac{\partial\Psi}{\partial t} = \left[ -\frac{\hbar^2}{2m} \nabla^2 + V_\text{trap} + \frac{4\pi\hbar^2a}{m} \left|\Psi\right|^2 - i\frac{\hbar}{2} K_3 \left|\Psi\right|^4 \right] \Psi \,,
\label{eqn:gpe}
\end{equation}
where $V_\text{trap}$ is the confining potential. Three-body recombination is modelled by the phenomenological inclusion of an imaginary loss term proportional to the three-body loss rate coefficient $K_3$ (which differs by a Bose statistical factor of $3!$ from the coefficient for noncondensed atoms). This term leads to loss proportional to the cube of the atomic density:
\begin{equation}
\frac{\partial}{\partial t} \int |\Psi|^2 d\vec{r} = -\int K_3 |\Psi|^6 d\vec{r} \,,
\end{equation}
and entails the assumption that the products of recombination collisions leave the trap without interacting with the remaining atoms. As three-body processes dominate at the high densities relevant to this experiment \cite{Saito:2002}, we do not include the effect of two-body inelastic collisions. To make the computation tractable, in integrating equation (\ref{eqn:gpe}) we assume a cylindrically symmetric trap with oscillation frequencies $\omega_{z,\rho} = 2\pi\times\{53,24\}$\,Hz, such that the mean trap frequency $\bar\omega$ matches that of our crossed dipole trap (the collapse time has been found to be relatively robust to asymmetry in the trapping potential \cite{Savage:2003}). The simulation includes the 100\,ms magnetic field ramp from $a=+89.1a_0$ to $a_\text{init}=+0.2a_0$, but we neglect the expansion of the condensate in our simulation, as the density spikes which trigger the recombination losses cease once the interactions are made repulsive.

The results of this simulation for $N_0 = 4\times10^4$ and $a_\text{init} = +0.2a_0$ are overlaid with the experimental data in Figure \ref{fig:collapsetimes} (solid line). For these simulations the three-body loss coefficient was scaled with $a_\text{collapse}$ as $K_3 = 8\times10^{-14}\,a^2$ cm$^4$/s \cite{Bao:2004}. The dotted lines show the variation in the simulated collapse time due to the combined experimental uncertainties in the initial scattering length and the trap frequencies, as well as run-to-run number fluctuations of 20\%. The simulations show good quantitative agreement with the experimental data.

It should be noted that the 100\,ms ramp of the scattering length from $a=+89.1a_0$ to $a=+0.2a_0$ is not truly adiabatic. Our GP simulations show that the ramp excites breathing mode oscillations, despite the duration of the ramp exceeding the mean trap oscillation period by a factor of 3. The oscillation is predominantly along the weak trapping axes, and has an amplitude of approximately 10\% of the condensate radius. It occurs because, although $a$ is varied smoothly, the condensate size does not depend linearly on the scattering length -- in the Thomas-Fermi limit, the radius scales as $a^{1/5}$. This excitation accelerates the contraction of the condensate, decreasing $t_\text{collapse}$ by approximately 15\%. The effect is included in the simulations shown in Figure \ref{fig:collapsetimes}. It could be reduced by tailoring the magnetic field ramp to ensure that the condensate radius decreases smoothly. Our simulations show that ramping $a^{1/4}$ smoothly over 100\,ms reduces the breathing mode oscillations to below 1\%, and causes the collapse times to be indistinguishable from those for a condensate that is in the ground state immediately prior to the collapse.

We now turn our attention to the possible systematics which may affect the agreement between theory and experiment. The source of the largest experimental uncertainty in our system is the oscillation frequencies of the crossed dipole trap, which vary during the evaporation to BEC as the intensity of the trapping laser is reduced. For technical reasons, we cannot directly measure the trap frequencies at the end of the evaporation. Instead, we make several measurements at higher intensities, which we fit to an analytic model of the dipole potential including the effect of gravity in the vertical direction. The model is further constrained by knowledge of the intensity $I_0$ at which gravity overcomes the dipole potential and the trap vanishes. Due to the strong dependence of the vertical trap frequency on the laser intensity near $I_0$, and the variation in the intensity itself, we estimate an uncertainty in $\bar\omega$ of 10\%, predominantly in the vertical direction. As the peak density of a noninteracting condensate scales as $\bar\omega^{3/4}$, this corresponds to an estimated uncertainty in $n_0$ of 7\%, which in turn produces an uncertainty in the simulated collapse time of approximately 15\%. This is not sufficient to explain the inconsistency between our results and the JILA experiment.

We must also consider the possibility of a systematic error in our determination of atom number. $N_0$ is calculated using the theoretical optical cross-section by integrating the optical depth of an absorption image. We image the atoms on resonance with circularly polarized light and apply a small bias magnetic field along the imaging direction to provide a quantization axis. The calculation therefore makes use of the resonant cross-section and saturation intensity of the cycling transition $\hf{F=3}{m_F=\pm3}\rightarrow\hf{F^\prime=4}{m_{F^\prime}=\pm4}$. As a result, our measured value of $N_0$ is a lower bound: any errors in the polarization or detuning of the imaging light, or in the alignment of the quantization field, will reduce the measured atom number. We estimate the uncertainty in $N_0$ due to these effects to be less than 5\%. Furthermore, if the atom number were undercounted then correcting for this would decrease the simulated collapse times, as a higher initial density speeds up the contraction. This effect therefore also cannot explain the disparity between our results and the original experiment, for which GP simulations \emph{over}estimated the collapse times.

Although $t_\text{collapse}$ is only weakly dependent on the three-body loss coefficient $K_3$, the shape of the loss curves is affected by varying this parameter. The values of $K_3$ used in simulations of the original JILA experiment ranged from $K_3 = 2\times10^{-28}$ cm$^6$/s \cite{Saito:2002} to $K_3 = 2\times10^{-26}$ cm$^6$/s \cite{Savage:2003}. Several authors also considered a relationship between the loss coefficient and the scattering length of the form $K_3 \sim |a|^2$ for $a<0$ \cite{Moerdijk:1996,Adhikari:2004}, with Bao \etal\ deducing $K_3 = 2.68\times10^{-13}\,a^2$ cm$^4$/s \cite{Bao:2004}. We find that these values cannot reproduce the shape of our measured loss curves.

Figure \ref{fig:k3collapse} shows the results of GP simulations using values of $K_3$ between $5\times10^{-27}$ cm$^6$/s and $5\times10^{-29}$ cm$^6$/s overlaid with experimental data for $a_\text{collapse}=-8.4a_0$. At higher loss rates, the high initial density of our sample causes significant loss during the contraction of the condensate in the simulation during the moments leading up to the collapse. In fact, for $K_3 > 10^{-27}$ cm$^6$/s this initial loss is so great that there is no sudden implosion of the condensate and no discernible elbow in the loss curve \footnote{To aid comparison of the elbow in the loss curve, the trap frequencies in the simulations have been adjusted to give the observed collapse time in Figure \ref{fig:k3collapse}. Nonetheless, as can be seen from Figure \ref{fig:collapsetimes}, the experimental collapse time at $a_\text{collapse}=-8.4a_0$ is slightly below -- although within the error bars of -- the simulated collapse time.}. In order to obtain the abrupt onset of loss that we observe in the experiment, a three-body loss rate of $K_3 \leq 5\times10^{-29}$ cm$^6$/s at $a=-8.4a_0$ is required. From a similar analysis of the $a_\text{collapse}=-20a_0$ data shown in Figure \ref{fig:collapse}(b) we find $K_3 \leq 1\times10^{-28}$ cm$^6$/s at that scattering length. These limits are more than an order of magnitude below most of the values used to simulate the original experiment. Assuming a scaling with $|a|^2$, they imply $K_3 \lesssim 1\times10^{-14} \, a^2$ cm$^4$/s. In this regime, loss after the initial implosion is caused by intermittent local density spikes between which three-body loss is negligible. This was first predicted by Saito and Ueda \cite{Saito:2001} even before the JILA experiment. These discrete implosions result in the numerous plateaus apparent in the simulated loss curve, although the scatter in our experimental data -- caused primarily by run-to-run fluctuations in atom number -- is too large to observe these directly.

\fig[t]{0.47}{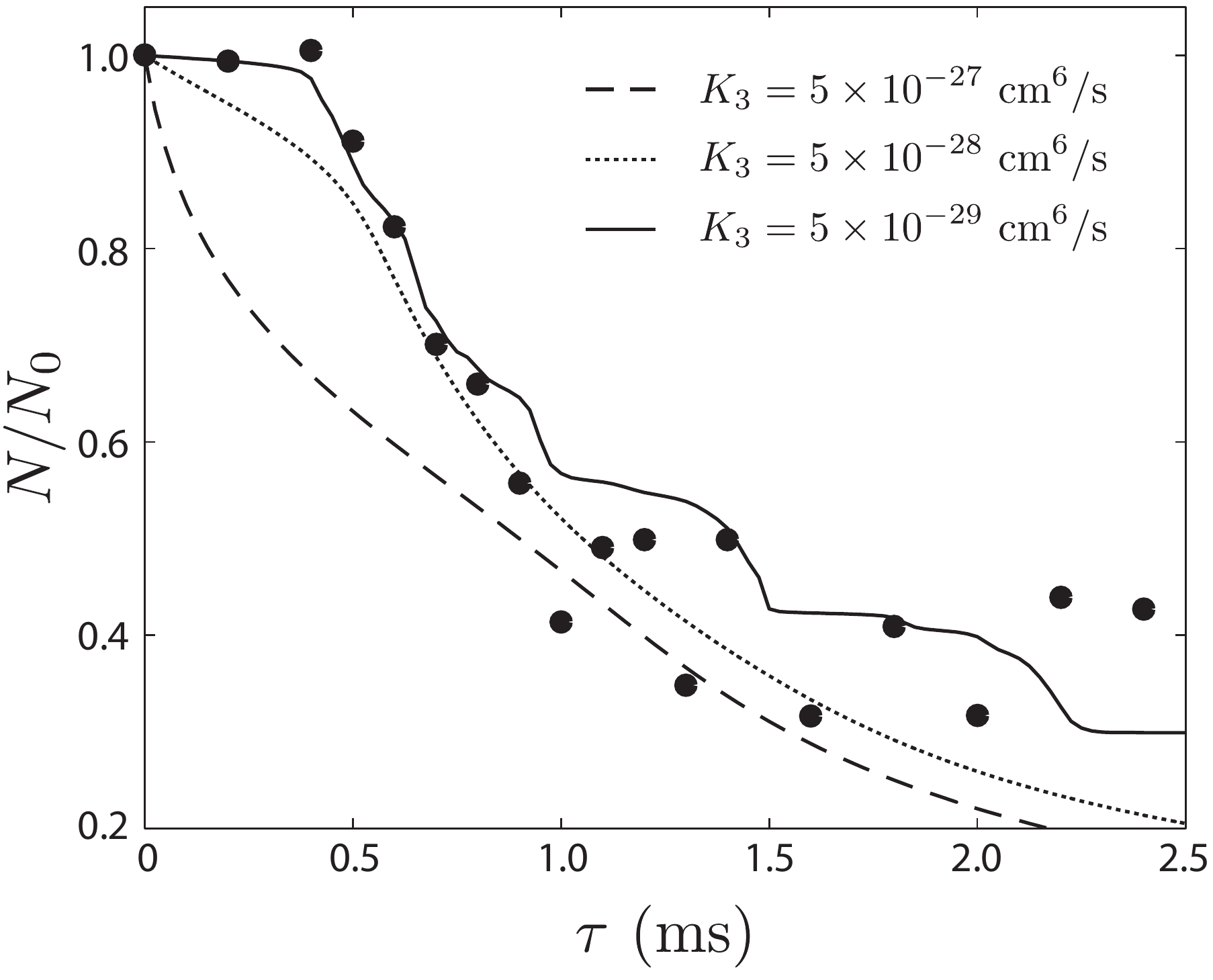}{Comparison of experimental and simulated collapse data for $a_\text{collapse}=-8.4a_0$. The data points show the measured atom number $N$ (normalized to $N_0$) as a function of evolution time $\tau$ at $a<0$, and the lines represent GP simulation results with different values of the three-body loss coefficient $K_3$. A value of $K_3 \leq 5\times10^{-29}$ cm$^6$/s is necessary to replicate the sudden onset of loss detected in the experiment.}

We have investigated the inelastic loss rates further by measuring the depletion of our condensates over time with positive scattering lengths, at which the condensates are stable. The rate at which atoms are lost due to two- and three-body inelastic collisions depends on the density profile of the condensate. In the limit that $a\rightarrow0$, the density is given by the modulus squared of the ground state harmonic oscillator wavefunction, and the loss rate equation $\dot{N}/N =-\sum_i K_i \langle n^{i-1} \rangle$ becomes:
\begin{eqnarray}
\dot{N} = -N/\tau - \eta_2 K_2 N^2  - \eta_3 K_3 N^3 \,,
\label{eqn:losseqnNI}
\end{eqnarray}
where $\tau$ represents the one-body loss rate, $\eta_2 = (2\pi a_\text{ho}^2)^{-3/2}$ and $\eta_3 = (\sqrt{3}\pi a_\text{ho}^2)^{-3}$. In the Thomas-Fermi limit $Na/a_\text{ho} \gg 1$, the condensate density takes on the shape of the confining potential and the loss rate equation evaluates to:
\begin{eqnarray}
\dot{N} = -N/\tau - \gamma_2 K_2 N^{7/5}  - \gamma_3 K_3 N^{9/5} \,,
\label{eqn:losseqnTF}
\end{eqnarray}
with $\gamma_2 = 15^{2/5}/[14\pi a^{3/5}a_\text{ho}^{12/5}]$ and $\gamma_3 = 5^{4/5}/[56\pi^2 3^{1/5} a^{6/5}a_\text{ho}^{24/5}]$. It should be noted that these expressions are valid only when the loss rate is small compared with the trap frequencies $K_i \langle n^{i-1} \rangle \ll \bar\omega$, so that the atomic density profile does not change significantly.

Figure \ref{fig:losscurves} shows the number of atoms remaining as a function of time in condensates with $a=0$ and $a=+37.6a_0$. The lines plot the best-fit solutions to (\ref{eqn:losseqnNI}) and (\ref{eqn:losseqnTF}) respectively, assuming that the loss is entirely due to two-body (solid) or three-body (dashed) inelastic collisions. It is difficult to separate the contributions of two- and three-body loss purely from the shape of the decay curve, as has been noted in previous work \cite{Roberts:2000}. Nonetheless, attributing all of the measured loss to two- or three-body processes allows us to place an upper limit on the value of $K_2$ and $K_3$ at these scattering lengths. Figure \ref{fig:lossrates}(a) shows these upper bounds for scattering lengths between $0$ and $+100a_0$. In our system, $Na/a_\text{ho} \simeq a/a_0$ and we use the Thomas-Fermi approximation (\ref{eqn:losseqnTF}) except at $a=0$. The error bars represent the statistical uncertainties in the fits; we assign an additional systematic error of 10\% to incorporate the uncertainty in $\bar\omega$.

\fig[t]{0.265}{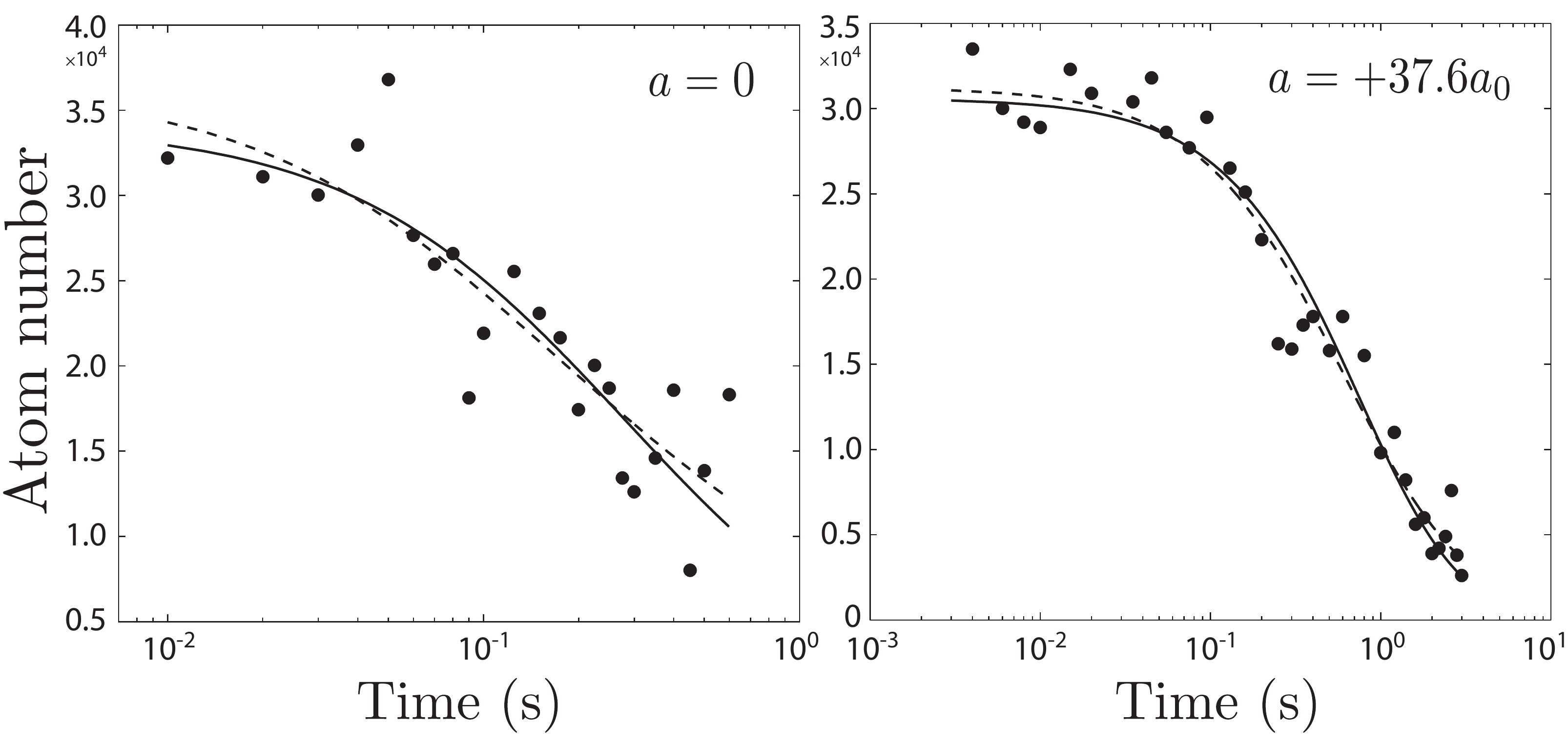}{Measurements of inelastic losses in \Rb{85} condensates. The data points show the atom number as a function of hold time in the optical trap for condensates with $a=0$ and $a=+37.6a_0$. The solid lines are fits of the solutions of (\ref{eqn:losseqnNI}) and (\ref{eqn:losseqnTF}) to the experimental data, assuming $K_3=0$ (solid) and $K_2=0$ (dashed). Although the contributions of two- and three-body processes cannot be distinguished in this manner, these fits may be used to place upper bounds on the values of $K_2$ and $K_3$.}

\fig[b]{0.41}{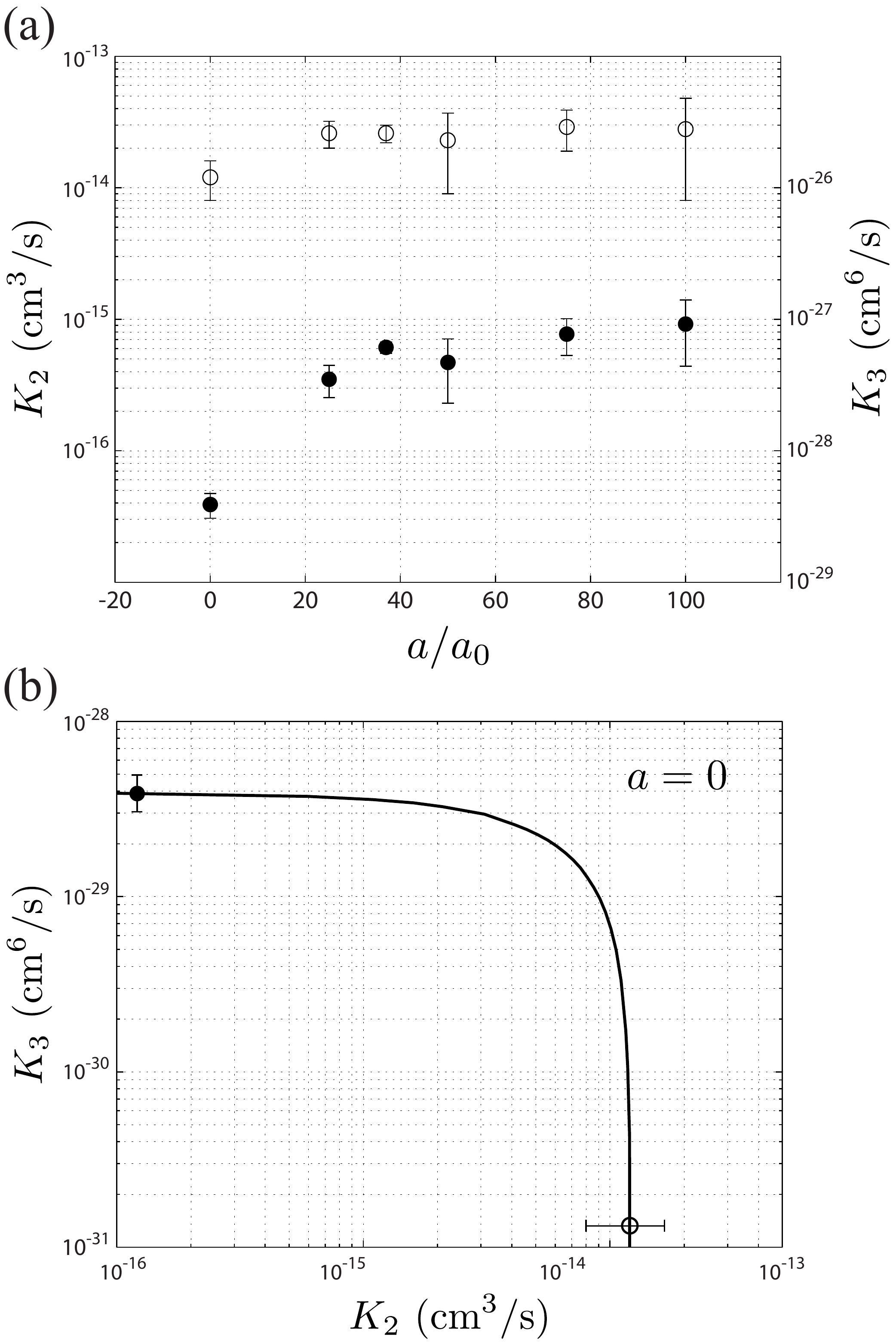}{(a) Upper bounds on $K_2$ (open circles) and $K_3$ (filled circles) as a function of scattering length, calculated from fits to the solutions of (\ref{eqn:losseqnNI}) and (\ref{eqn:losseqnTF}). The error bars represent statistical uncertainties. (b) Locus of two- and three-body loss coefficients for which the solution to (\ref{eqn:losseqnNI}) fits the experimental data for $a=0$ shown in Figure \ref{fig:losscurves}. The $x$ and $y$ intercepts correspond to the upper bounds shown in (a). Assuming $K_2 \simeq 1.2\times10^{-14}$ cm$^3$/s \cite{Roberts:2000}, the data suggest a three-body loss coefficient $K_3 \leq 10^{-30}$ cm$^6$/s.}

Theoretical calculations suggest that the recombination rate should vary strongly with the two-body elastic scattering cross-section, with several authors predicting a universal $K_3 \sim a^4$ scaling in the zero-temperature limit \cite{Fedichev:1996,Nielsen:1999,Esry:1999}. Our observations are consistent with a strong suppression of the recombination rate at the zero crossing of the $s$-wave scattering length, with the measured upper bound $K_3 \leq (3.9\pm0.7)\times10^{-29}$ cm$^6$/s at $a=0$ an order of magnitude below that for $a>+50a_0$, and more than three orders of magnitude below the loss rate far from the Feshbach resonance, $K_3 = 7\times10^{-26}$ cm$^6$/s \cite{Roberts:2000}.

We can combine this latest data with previous measurements of the two-body loss rate to further constrain the three-body recombination coefficient. Figure \ref{fig:lossrates}(b) shows the locus of possible $K_2,K_3$ values for which the solution to (\ref{eqn:losseqnNI}) best fits our experimental loss curve at $a=0$. In Ref. \cite{Roberts:2000}, Roberts \etal\ measured $K^\text{nc}_2 \simeq 2.4\times10^{-14}$ cm$^3$/s for thermal clouds in the vicinity of $a=0$, corresponding to a value of $K_2 \simeq 1.2\times10^{-14}$ cm$^3$/s for condensed atoms. This matches our measured upper bound of $K_2 \leq (1.2\pm0.2)\times10^{-14}$ cm$^3$/s. Coupled with this result, our data is consistent with a value of the three-body loss coefficient $K_3 \leq 10^{-30}$ cm$^6$/s. Ref. \cite{Esry:1999} predicts $K_3 \simeq 5\times10^{-32}$ cm$^6$/s at $a=0$. In comparison, the three-body loss coefficient for \Rb{87} is $K_3 = 6\times10^{-30}$ cm$^6$/s \cite{Burt:1997}.

In conclusion, we have presented new experimental data on the collapse of \Rb{85} Bose-Einstein condensates with attractive interactions in an optical dipole trap. Our results qualitatively match those of the original JILA bosenova experiment, but in addition agree quantitatively with GP simulations. We find that a lower value of the three-body loss coefficient $K_3$ than was used in simulating the original experiment is needed to reproduce the sudden onset of loss that we observe. We have also analysed the decay of atoms from our condensates and thereby placed further constraints on the three-body loss coefficient $K_3$ at small positive scattering lengths. We expect that this work will inform future experimental and theoretical investigations of this rich quantum system.

We thank J. Hope, M. Johnsson, S. Szigeti and M. Hush for helpful discussions. This work was supported by the Australian Research Council Centre of Excellence for Quantum-Atom Optics.

\bibliography{bsnv}

\end{document}